\documentclass[aps,prl,preprint,groupedaddress,showpacs]{revtex4}

\usepackage{graphics}
\usepackage{bm}

\usepackage{amsmath}    
\usepackage{amssymb}    
\usepackage{graphicx}   
\usepackage{epsf}
\usepackage{epsfig}
\begin{document}

\title{Spontaneous Fluxoid Formation in Superconducting Loops}
\thanks{Submitted to Phys. Rev. Letts.}

\author{R. Monaco}
\affiliation{Istituto di Cibernetica del CNR, 80078, Pozzuoli, Italy
and Unit$\grave{\rm a}$ INFM $­$ Dipartimento di Fisica, Universit$\grave{\rm a}$ di Salerno, 84081 Baronissi, Italy}\email{roberto@sa.infn.it}
\author{J. Mygind}
\affiliation{DTU Physics, B309, Technical University of
Denmark, DK-2800 Lyngby, Denmark}\email{myg@fysik.dtu.dk}
\author{R.\ J.\ Rivers}
\affiliation{Blackett Laboratory, Imperial College London, London
SW7 2AZ, U.K. }\email{r.rivers@imperial.ac.uk}
\author{V.\ P.\ Koshelets}
\affiliation{Kotel'nikov Institute of Radio Engineering and Electronics Russian Academy of Science, Mokhovaya 11, Bldg 7, 125009, Moscow, Russia.}\email{valery@hitech.cplire.ru}

\date{\today}

\begin{abstract}
We report on the first experimental verification of the Zurek-Kibble
scenario in an isolated superconducting ring over a wide parameter
range. The probability of creating a single flux quantum
spontaneously during the fast normal-superconducting phase
transition of a wide Nb loop clearly follows an allometric
dependence on the quenching time $\tau _{Q}$, as one would expect if
the transition took place as fast as causality permits. However, the
observed Zurek-Kibble scaling exponent $\sigma = 0.62\pm0.15$ is two
times larger than anticipated for large loops. Assuming Gaussian
winding number densities we show that this doubling is well-founded
for small annuli.

\end{abstract}

\maketitle

The Zurek-Kibble (ZK) scenario \cite{zurek1,zurek2,kibble1} proposes
that continuous phase transitions take effect as fast as possible
i.e. the domain structure after the quenching of the system
initially reflects the causal horizons. This proposal can be tested
directly for transitions whose domain boundaries carry visible
topological charge. In this letter we shall show that, with
qualifications, this scenario is strongly corroborated by the
behaviour of superconducting loops, for which the topological defect
is a fluxoid i.e. a supercurrent vortex carrying one magnetic flux
quantum $\Phi_0 = h/(2e)\simeq 2.08 \times 10^{-15}\,$Wb.

The basic scenario is very simple. Consider a planar low-$T_c$
superconductor in which a hole has been made of circumference $C$.
In the Meissner state the order parameter for the superconductor is
a complex field $\psi$ with phase $\phi$, $\psi = \rho e^{i \phi}$,
where $|\rho|^2$ measures the density of Cooper pairs. On quenching
the system from the normal to superconducting phase, causality
prevents the system from adopting a uniform phase. If, on completion
of the quench, we follow the periodic phase $\phi(x)$ (mod $2\pi$)
along the boundary of the hole (co-ordinate $ x$), we can define a
winding number density: $ n(x) = d\phi(x)/dx/(2\pi)$. The
total normalized magnetic flux through the hole is, in units of
$\Phi_0$, the winding number:

\begin{equation}
n =\int_{0}^{C}\, n\left(x\right) dx = \frac{\Delta\phi}{2\pi},
\label{intn}
\end{equation}

\noindent where $\Delta\phi$ is the change in $\phi$. In the absence
of an external magnetic field, on average $\langle n\rangle = 0$,
but it will have non-zero variance $(\Delta n)^2=\langle
n^2\rangle$, which is what can be measured in terms of the
probabilities $f_{\pm m}$ to trap $\pm$m flux quanta: $\langle
n^2\rangle= \sum_{m=-\infty}^\infty m^2 f_m$. According to Ref.\cite{zurek2}, on completion of a thermal quench having a given inverse quench rate
$\tau_{Q}=-T_{c}/(dT/dt)_{T=T_{c}}$, the phase $\phi$ is correlated
over distances $2\pi\bar\xi$, where $\bar\xi$ was predicted to
depend allometrically\cite{zurek2} on the quench time $\tau_Q$:

\begin{equation}
{\bar{\xi}}\approx {\xi _{0}}\bigg(\frac{\tau _{Q}}{\tau _{0}}\bigg)^{\sigma}.  \label{xi}
\end{equation}


\noindent  ${\bar\xi}$, also called the ZK causal length, is defined
in terms of the cold correlation length $\xi_0$ and the
Ginzburg-Landau relaxation time $\tau_0$ of the long wavelength
modes. The ZK scaling exponent $\sigma$ is determined by the {\it
static} critical exponents of the system and, in the mean-field
approximation, $\sigma=1/4$ \cite{zurek1}. If we make the further
assumption that there is a random walk in phase on a scale
$2\pi{\bar\xi}$ then, for a hole of radius $r$, circumference $C\gg
2\pi{\bar\xi}$,

\begin{equation}
\langle n^2\rangle \approx \frac{C}{2 \pi \bar{\xi}}=\frac{r}{\xi _{0}}\bigg(\frac{\tau _{Q}}{\tau _{0}}\bigg)^{-\sigma }.  \label{var}
\end{equation}

For small rings with $C < 2\pi{\bar\xi}$ the likelihood of seeing
two or more units of flux is small and $\langle n^2\rangle \approx
f_{+1} + f_{-1} =f_1$, the probability of single fluxoid trapping.
It is plausible to extrapolate Eq.(\ref{var}) to:

 \begin{equation}
f_1 \approx \langle n^2\rangle  \approx \frac{r}{\xi _{0}}\bigg(\frac{\tau _{Q}}{\tau _{0}}\bigg)^{-\sigma },  \label{varsmall}
\end{equation}

\noindent showing allometric behaviour of $f_1$ with the {\it same}
exponent. We note that the ZK argument makes no assumptions about
the rest of the superconductor, equally valid for the phase change
along the inner circumference of an annulus as it is for the phase
change around a single hole in a superconducting sheet. In 2003 the
first experiment with superconducting loops \cite{Kirtley} was
performed to test Eq.(\ref{var}). The experiment consisted of taking
an isolated array of thin-film wide rings and making it undergo a
forced phase transition by heating it above its superconducting
critical temperature and letting it to cool passively back towards
the $LHe$ temperature. Once the thermal cycle is over, the rings are
inspected by a scanning SQUID and the number and polarity of any
trapped fluxoids determined. These rings, of amorphous Mo$_3$Si thin
films, had thickness almost one order of magnitude smaller than the
low temperature London penetration depth. Although this provides
favorable conditions for thermally activated phenomena it
drastically increases the likelihood that nucleated vortices escape
through the ring walls during the fast quench. In fact, the
experimental outcome was totally at variance with the allometric
scaling above. However, the prediction Eq.(\ref{var}) presupposes
that we can ignore the contribution to the flux from the freezing in
of thermal fluctuations of the magnetic field \cite{rajantie} and
the results of \cite{Kirtley} could be explained in terms of the
freezing of thermally activated fluxoids in a similar spirit to
Ref.\cite{rajantie}.

\smallskip In this paper we shall present results from a new
experiment with high-quality Nb film rings with $r=30\,\mu$m, two
times thicker than their low temperature London penetration length
$\lambda_{L,Nb}$; the film thickness and composition were chosen to
reduce the thermal activation of fluxoids and, at the same time,
'the washing out' of fluxoids generated by the conventional
causality mechanism. In our case the contribution $\Delta f_1$ to
the probability of finding a unit of flux from thermal magnetic
field fluctuations is approximately\cite{rajantie}:

\begin{equation}
\Delta f_1 \lesssim (k_B T_c)r\mu_0/\Phi_0^2 \approx 6\times
10^{-4},
\end{equation}

\noindent and can be safely ignored. We therefore look for scaling
behaviour in $\tau_Q$.

\begin{figure}[tb]
        \centering
                \includegraphics[width=9cm]{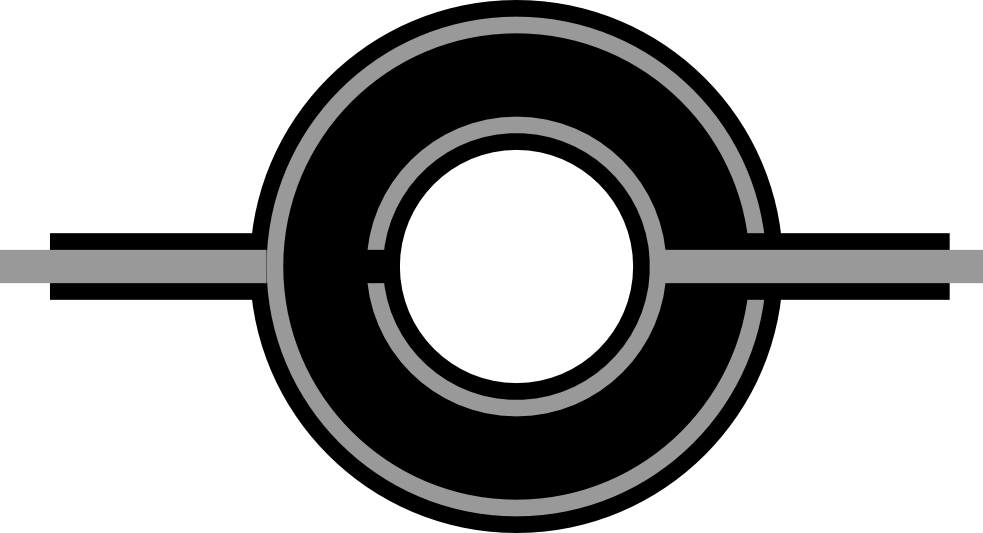}
        \caption{Sketch of a superconducting loop (black) used as a base electrode for two gapped Nb/AlOx/Nb annular Josephson tunnel junctions (whose top electrodes are in gray). The ring inner and outer radii are $r=30$ and $R=50 \mu$m, respectively, while the top electrodes width is $5 \mu$m.}
        \label{Geometry}
\end{figure}

Here a different way of counting both the number and the polarity of
generated defects has been adopted. It is based on the detection of
the persistent currents $\bf{J_s}$ circulating around a hole in a
superconducting film, when one or more fluxoids are trapped inside
the hole. The circulating currents screening the bulk of the
superconductor from the trapped flux induce a magnetic field
$\bf{H}$ in the volume around the ring, such that
$\bf{J_s}=\bf{\nabla} x \bf{H}$. By placing a Josephson tunnel
junction (JTJ) along the perimeter of the hole in the area where
this field passes, any trapped fluxoid will result in a modulation
of the JTJ critical current, similar to the effect of an external
field applied perpendicular to the ring. Indeed, this method is
strongly inspired by the results found investigating the effects of
a transverse field on Josephson junctions of various
geometries\cite{JAP08}. The geometry of our experiment is sketched
in Fig.\ref{Geometry}; the black wide ring is a $200\,$nm thick Nb
film, which also acts as the common base electrode for two JTJs
whose top electrodes are depicted in grey. The JTJs have the shape
of gapped annuli and the bias current is supplied in their middle
point: to our knowledge, this geometrical configuration has never
been realized before and is characterized by a peculiar magnetic
diffraction pattern: for small magnetic fields the critical current
increases both for positive and negative field values. The original
purpose of having two counter electrodes on the base ring was that
any screening current circulating on the outer ring circumference
will preferentially affect the outermost JTJ, and vice versa for the
screening current on the inside of the ring. Since the persistent
currents due to trapped flux mainly flow in the inner ring
circumference \cite{clem}, in this experiment we only used the
innermost JTJ. The layout shown in Fig.\ref{Geometry}, with a few
key differences, bears remarkable topological similarity to the one
used in a series of experiments by us to demonstrate the ZK scaling
behaviour of Eq.(\ref{var}) in annular JTJs
\cite{PRL01,PRL02,PRB03,PRL06,PRB08}. The most obvious difference in
the design is the inclusion of the two junction counter electrodes
on top of the ring-shaped base electrode. The second change is the
removal of a small section of the full annular junction to leave a
gapped annular junction with the purpose of avoiding fluxons created
inside the JTJ at the Josephson phase transition. Should any be
produced, they will simply migrate through the junction extremities
driven by the applied bias current needed to overcome eventual
pinning potentials. This leaves the experiment only sensitive to the
fluxoids produced in the ring at the phase transition.

\noindent As with our previous experiments, the present one relies
on a fast heating system, obtained by integrating a Mo resistive
meander line on the $4.2$mm$\times 3$mm$\times 0.35$mm Si chip
containing the ring with the Nb/AlOx/Nb JTJs. The quench time $\tau
_{Q}$ could be continuously varied over more that four orders of
magnitude (from $20\,$s down to $1\,$ms) by varying the width and
the amplitude of the voltage pulse across the integrated resistive
element. In order to determine the quench time with high accuracy,
the ring temperature was monitored exploiting the well known
temperature dependence of the gap voltage of high-quality Nb/AlOx/Nb
JTJs already described in Ref.\cite{PRB03}. After each ring thermal
quench the critical current of the innermost JTJ is automatically
stored and an algorithm has been developed for the counting of the
trapped fluxoids. Finally, all the measurements have been carried
out in a magnetic and electromagnetically shielded environment.
During the thermal quenches all electrical connections to the heater
as well as to the JTJs were disconnected. While more details on the
measurement setup and on the fabrication process can be found in
Ref.\cite{PRB09} and Ref.\cite{VPK}, respectively, an extensive
description of the chip layout, the experimental setup and the
system calibration will be given elsewhere\cite{next2}.

\begin{figure}[tb]
\centering
\includegraphics[width=9cm]{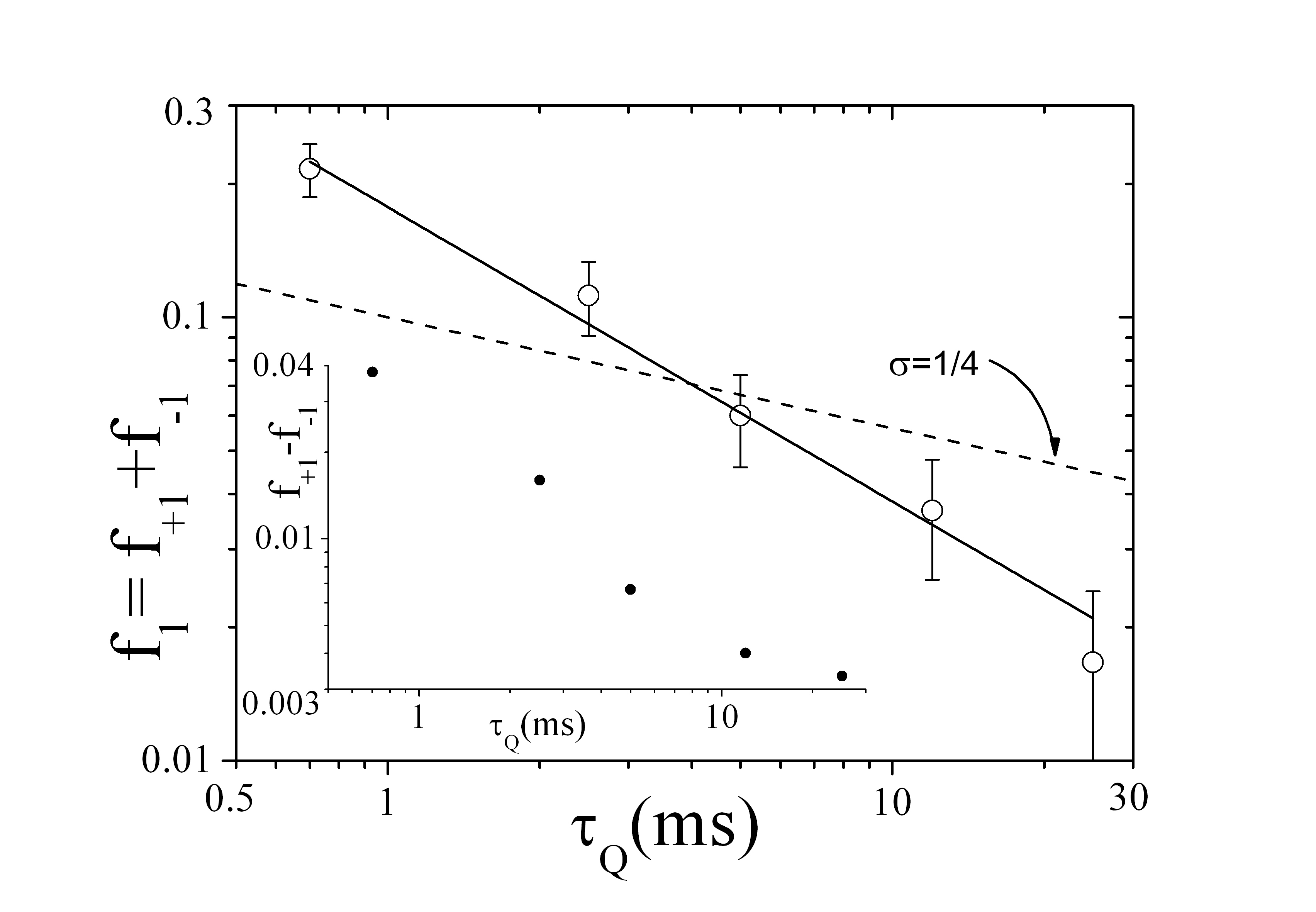}
\caption{ Log-log plot of the measured frequency $f_{1}$ of trapping single fluxoid versus the quenching time $\tau _{Q}$ for a Nb ring having inner radius $r=30\,\mu$m, outer radius $R=50\,\mu$m and thickness $d=200\,$nm $\simeq 2 \lambda_{L,Nb}$. Each point corresponds to hundreds of thermal cycles. The vertical error bars gives the statistical error, while the relative error bars in $\tau _{Q}$ amounting to $\pm10\%$ are as large as the dots' width. The solid line is the best fit to an allometric relationship $f_{1}=a\,\tau_{Q}^{-b}$ which yields  $a=0.18\pm 0.02$ (taking $\tau_{Q}$ in ms) and $b=0.62\pm 0.15$. For comparison purposes, the dashed line is the prediction of Eq.(\ref{g3})) with $\chi$ (see text) set to $0.7$ to fit in ordinate scale.}
\label{Fig.2}
\end{figure}

\smallskip The experimental results shown in Fig.~\ref{Fig.2} were
obtained using a ring with inner and outer radii, $r=30\,\mu$m and
$R=50\,\mu$m, respectively. Similar samples have shown the same
behaviour. We note that wider rings prevent fluxoids from tunneling
out of the ring, although their smaller normal self-inductance $L_n$
makes  fluxoid formation energetically more unlikely. In our case,
the field energy $E_0=\Phi_0^2/2L_n$ associated with a single flux
quantum $\Phi_0$ is several orders of magnitude larger than the
thermal energy $k_B T_c/2$ at $T_c$\cite{note}.

Magnetostatic numerical simulations implemented in the COMSOL
Multiphysics 3D Electromagnetics module showed that, when  a single
flux quantum is trapped in such rings, the radial magnetic field
induced by the circulating currents at the ring inner border is as
large as $1$A/m, a value easily detectable by the JTJ. 
For our rings, the number of trapped fluxoids was small, usually no
more than one; indeed we measured the probability of trapping a
single up-fluxoid (field up) $f_{+1}$ and a single down-fluxoid $f_{-1}$.

Fig.~\ref{Fig.2} shows on a log-log plot the measured frequency $
f_1 = f_{+1}+f_{-1}= (n_{+1}+n_{-1})/N=n_{1}/N$ of single fluxoid
trapping, obtained by quenching the sample $N$ times for each value
of a given quenching time $\tau _{Q}$, $n_{1}$ being the number of
times that one defect or one anti-defect was spontaneously produced.
$N$ ranged between $250$ and $300$ and $n_{1}$ was never smaller
then $10$, except for the rightmost point for which $n_{1}=5$. The
vertical error bars gives the statistical error $f_{1}/\surd n_1$.
The relative error bars in $\tau _{Q}$ amounting to $\pm10\%$ are as
large as the dot's width. As expected we had $n_{+1}\approx n_{-1}$,
but slightly larger than $n_{-1}$ indicating the presence of a small
residual stray field in our apparatus (see inset of Fig.2).

\noindent To test Eq.(\ref{xi}), we have fitted the data with an
allometric function $f_{1}=a\,\tau _{Q}^{-b}$, with $a$ and $b$ as
free fitting parameters. An instrumentally weighted
least-mean-square fit of $f_{1}$ vs. $\tau _{Q}$, represented by the
continuous line in Fig.~\ref{Fig.2}, yields $a=0.18 \pm 0.02$
(taking $\tau _{Q}$ in ms) and $b=0.62\pm 0.15$. The large fit
correlation coefficient $R^2=0.987$ indicates that the allometric
behaviour is reliably confirmed, however the scaling exponent $b$ is
about two times larger than expected for large loops.

A {\it doubling} of the large-loop ZK exponent for small loops has a
possible explanation in the framework of the Gaussian correlation
model introduced in Ref.\cite{PRB08} in which it was assumed that
the winding number $n(x)$ is a {\it Gaussian} variable until the
transition is complete, whereby all correlation functions are
determined by the two-point correlation function $g(x_1-x_2, C)
=\left\langle n\left(x_1\right)n\left(x_2\right)\right\rangle$. As a
result \cite{ray01}:

\begin{equation}
\langle n^2\rangle =\int_{0}^{C} \int_{0}^{C} \left\langle
n\left(x_1\right)n\left(x_2\right)\right\rangle dx_1dx_2
=2C \int_{0}^{C} g(x,C)dx.
\label{g2}
\end{equation}

\noindent For $C \ll 2\pi{\bar\xi}$, we can assume a correlation
function of the form $g(x, C) = {\bar g}(x/{2\pi\bar\xi},
C/{2\pi\bar\xi})/({2\pi\bar\xi})^2$, so that:

\begin{equation}
f_1\approx 2\frac{C}{2\pi\bar\xi} \int_{0}^{C/2\pi{\bar\xi}} {\bar
g}({\bar x}, C/2\pi{\bar\xi})d{\bar x} \approx 2\bigg(\frac{C}{2\pi\bar\xi}\bigg)^2 {\bar g}(0, C/2\pi{\bar\xi}).
\label{g4}
\end{equation}

\noindent {\it provided} ${\bar g}({\bar x}, C/2\pi{\bar\xi})$ is
analytic at ${\bar x}=0$. This suggests that Eq.(\ref{varsmall})
should be replaced by the scaling behaviour:

\begin{equation}
 f_1\approx \kappa \bigg(\frac{C}{2 \pi \bar{\xi}}\bigg)^2= \kappa\bigg(\frac{r}{\xi _{0}}\bigg)^2\bigg(\frac{\tau _{Q}}{\tau _{0}}\bigg)^{-2\sigma },  \label{var2}
\end{equation} 

\noindent with a proportionality constant $\kappa$ of the order of
unity. To buttress this suggestion, it is not difficult to show
that, in the Gaussian approximation, the value of $\langle
n^2\rangle$ along a small ring in a 2D superconductor is
proportional to the area enclosed by the ring \cite{next2}.

\noindent This doubling of the scaling exponent in Eq.(\ref{var2})
has the price of coming with a lower probability, but leaves us with
some freedom with the ZK prefactor. Indeed, the value of the
prefactor $a$ obtained from the allometric best fit of the
experimental data in Fig.\ref{Fig.2} is about $\kappa=4$-$5$ times
larger than the predicted value $(r/\xi_{0})^2 \sqrt \tau_0 = 0.04$
obtained using the values $r=30\, \mu$m, $\xi _{0}\approx 30\,$nm
and $\tau_0=\pi \hbar/16 k_B T_c \approx 0.16\,$ps and taking
$\tau_{Q}$ in ms. As a bound we only expect agreement in the overall
normalization of the prefactor $a$ to somewhat better than an order
of magnitude, largely confirmed by experiment. We point out that the
dependence of the prefactor $a$ on the ring width remains to be
investigated both theoretically and experimentally. [We note that,
if Eq.(\ref{var}) were true, then  $\langle n^2\rangle$ would be
$\approx 0.6$, i.e., $20$ times larger than the experimental value
for $\tau_Q=O(10ms)$, say.]

In the opposite case of {\it large} circumferences, $C\gg
{2\pi\bar\xi}$, $g(x,C)$ is controlled by the correlation length
$\bar\xi$ of the winding number at the time of unfreezing and does not
depend on $C$, i.e., the effect of periodicity for large rings is
small. With $g(x) = {\bar g}(x/2\pi{\bar\xi})/(2\pi{\bar\xi})^2$ on
dimensional grounds, we justify the random walk assumption of
Eq.(\ref{var}),

\begin{equation}
\langle n^2\rangle = \frac{C}{\pi\bar\xi} \int_{0}^{C/2\pi{\bar\xi}} {\bar
g}(z) dz \approx \frac{C}{\pi\bar\xi} \int_{0}^{\infty} {\bar g}(z) dz =
\chi \frac{C}{2\pi\bar\xi},
\label{g3}
\end{equation}

\noindent with $\chi = O(1)$.  The dashed line in Fig.\ref{Fig.2} is
the prediction in Eq.(\ref{g3}) with $\chi$ set to $0.7$ to fit to the
ordinate scale.

In summary, our experiment shows reliable scaling behaviour of the
form Eq.(\ref{var}) for the creation of a single fluxoid, with
scaling exponent $0.62 \pm 0.15$. This is obviously at variance with
the extrapolation (\ref{varsmall}) of the Zurek prediction
Eq.(\ref{var}) to small rings, for which we expect $\sigma = 0.25$.
We have suggested that it be given by Eq.(\ref{var2}) with twice the
exponent, as a consequence of the Gaussianity of the Cooper pair
field phase (before truncation by back-reaction), an assumption
supported in a slightly different context by the behaviour of JTJs
in an external field \cite{PRB08}. As Gaussianity permits the
instabilities from which defects form to grow as fast as possible,
in general it gives the same scaling exponents as the ZK scenario
for large systems. With this qualification we see our result as
providing strong support for Zurek-Kibble scaling over a wide range
of quenching time $\tau_Q$.  We stress that this experiment is the
\textit{only} one to date to have confirmed the Zurek-Kibble
causality scenario for single isolated superconducting rings (as
distinct from Josephson junctions). Further experiments to
investigate the transition to the random walk regime and the effect
of the ring width are planned. For example, a test of Gaussianity is
that $f_1\lesssim 0.5$ for all values of $C$ \cite{PRB08}.

\noindent Given that the original ZK scenario was posed to demonstrate the
similarity in the role of causality at transitions in the early
Universe and in condensed matter systems we have seen that the
finiteness of the latter systems requires careful disentangling from
the underlying principles before we can draw any quantitative
conclusions.

In the same vein we conclude with a speculation concerning the
small-annulus JTJ experiments \cite{PRL02,PRB03,PRL06,PRB08} that,
hitherto, have been the only superconductor experiments to show
scaling behaviour. In that case also, the observed exponent was
twice that anticipated from long annuli \cite{PRL01}. However, in
the case of JTJs there is an ambiguity in their fabrication that is
sufficient to double the exponent, according as the 'proximity
effect' enables otherwise subcritical behaviour of the Josephson
current density to dominate near the transition \cite{PRB03}. We had
assumed that this was the reason for the discrepancy. We shall now
reexamine these earlier experiments with the above analysis in mind.

\smallskip The authors thank P. Dmitriev for the sample fabrication and
testing, A. Gordeeva for useful discussions and M. Aaroe for the
help at the initial stage of the experiment.

\end{document}